\documentclass[aps,prb,reprint,superscriptaddress,showpacs]{revtex4-1}

\usepackage{amsmath,amssymb,graphicx}
\usepackage{epstopdf}
\usepackage{color}
\usepackage{MnSymbol}
\usepackage[version=3]{mhchem} 

\begin{document}


\title{Electroluminescence of monolayer \ce{WS_2} in a scanning tunneling microscope:  the effect of bias polarity on the spectral and angular distribution of the emitted light}



\author{Ricardo Javier Pe\~{n}a Rom\'{a}n}
\thanks{These authors contributed equally to this work.}
\affiliation{Institute of Physics ``Gleb Wataghin'', Department of Applied Physics, State University of Campinas-UNICAMP, 13083-859, Campinas, Brazil}

\author{Delphine Pommier}
\thanks{These authors contributed equally to this work.}
\affiliation{Universit\'{e} Paris-Saclay, CNRS, Institut des Sciences Mol\'{e}culaires d'Orsay, 91405, Orsay, France}

\author{R\'{e}mi Bretel}
\thanks{These authors contributed equally to this work.}
\affiliation{Universit\'{e} Paris-Saclay, CNRS, Institut des Sciences Mol\'{e}culaires d'Orsay, 91405, Orsay, France}

\author{Luis E. Parra L\'{o}pez}
\affiliation{Institut de Physique et de Chimie des Mat\'{e}riaux de Strasbourg, Universit\'{e} de Strasbourg, CNRS, IPCMS, UMR 7504, 67000 Strasbourg, France}

\author{Etienne Lorchat}
\affiliation{NTT Physics and Informatics Laboratories, NTT Research, Inc., 1950 University Ave., East Palo Alto, California 94303, USA.}

\author{Julien Chaste}
\affiliation{Universit\'{e} Paris-Saclay, CNRS, Centre de Nanosciences et de Nanotechnologies, 91120, Palaiseau, France}

\author{Abdelkarim Ouerghi}
\affiliation{Universit\'{e} Paris-Saclay, CNRS, Centre de Nanosciences et de Nanotechnologies, 91120, Palaiseau, France}

\author{S\'{e}verine Le Moal}
\affiliation{Universit\'{e} Paris-Saclay, CNRS, Institut des Sciences Mol\'{e}culaires d'Orsay, 91405, Orsay, France}

\author{Elizabeth Boer-Duchemin}
\affiliation{Universit\'{e} Paris-Saclay, CNRS, Institut des Sciences Mol\'{e}culaires d'Orsay, 91405, Orsay, France}

\author{G\'{e}rald Dujardin}
\affiliation{Universit\'{e} Paris-Saclay, CNRS, Institut des Sciences Mol\'{e}culaires d'Orsay, 91405, Orsay, France}

\author{Andrey G. Borisov}
\affiliation{Universit\'{e} Paris-Saclay, CNRS, Institut des Sciences Mol\'{e}culaires d'Orsay, 91405, Orsay, France}

\author{Luiz F. Zagonel}
\thanks{These authors contributed equally to this work.}
\affiliation{Institute of Physics ``Gleb Wataghin'', Department of Applied Physics, State University of Campinas-UNICAMP, 13083-859, Campinas, Brazil}

\author{Guillaume Schull}
\affiliation{Institut de Physique et de Chimie des Mat\'{e}riaux de Strasbourg, Universit\'{e} de Strasbourg, CNRS, IPCMS, UMR 7504, 67000 Strasbourg, France}

\author{St\'{e}phane Berciaud}
\affiliation{Institut de Physique et de Chimie des Mat\'{e}riaux de Strasbourg, Universit\'{e} de Strasbourg, CNRS, IPCMS, UMR 7504, 67000 Strasbourg, France}

\author{Eric Le Moal}
\email{eric.le-moal@universite-paris-saclay.fr}
\affiliation{Universit\'{e} Paris-Saclay, CNRS, Institut des Sciences Mol\'{e}culaires d'Orsay, 91405, Orsay, France}


\date{\today}

\begin{abstract}
 
Inelastic electron tunneling in a scanning tunneling microscope (STM) is used to generate excitons in monolayer tungsten disulfide (\ce{WS_2}). Excitonic electroluminescence is measured both at positive and negative sample bias. Using optical spectroscopy and Fourier-space optical microscopy, we show that the bias polarity of the tunnel junction determines the spectral and angular distribution of the emitted light. At positive sample bias, only emission from excitonic species featuring an in-plane transition dipole moment is detected. Based on the spectral distribution of the emitted light, we infer that the dominant contribution is from charged excitons, i.e., trions. At negative sample bias, additional contributions from lower-energy excitonic species are evidenced in the emission spectra and the angular distribution of the emitted light reveals a mixed character of in-plane and out-of-plane transition dipole moments. 
  
\end{abstract}


\maketitle


\section{Introduction \label{Intro}}

Monolayer transition-metal dichalocogenides (TMDs) are two-dimensional (2D) semiconductors that are actively considered for future device technologies~\cite{Mak2010,Splendiani2010,Mak2016,Brar2017,Wang2018}. Moreover, TMDs offer unique opportunities for fundamental research on exciton physics in 2D materials. Recently, scanning tunneling microscopy (STM)-induced luminescence (STML) has emerged as a promising nanoscopic probe of the optoelectronic properties of monolayer TMDs~\cite{Krane2016,Pommier2019,PenaRoman2020,Pechou2020,Schuler2020,ParaLopez2022}. There is experimental evidence that, in certain conditions, STML reveals the intrinsic excitonic species of the excited TMD monolayer~\cite{Pommier2019}. Nevertheless, the differences between STML and laser-induced photoluminescence (PL) of 2D semiconductors and the effect of the STM parameters on the STML of these materials are still open questions. In particular, much remains to be done to understand the dependence of such STML measurements on the bias polarity of the tip-sample junction.

In this article, we compare the STML and PL spectra of monolayer tungsten disulfide \ce{WS_2} and we report on the effect of the bias polarity on the STML spectrum and radiation pattern. All experiments are carried out under ambient conditions on a transparent conducting substrate and using a non-plasmonic tip, in order to avoid extrinsic effects due to surface plasmons at the emission frequencies of the excitons in the 2D semiconductor. Thus, no plasmon modes participate in the excitation process and the emitting excitons are not coupled to any plasmon mode of the tip, substrate or tip-sample gap. The STML is spatially and angularly resolved using an optical microscope. The angular distribution of STML reveals the orientation of the transition dipole moment of the emitters, which in general is key for establishing the excitonic nature of STML on 2D semiconductors~\cite{Pommier2019}.    

\section{Methods \label{Methd}}

%
\begin{figure*}
	\includegraphics{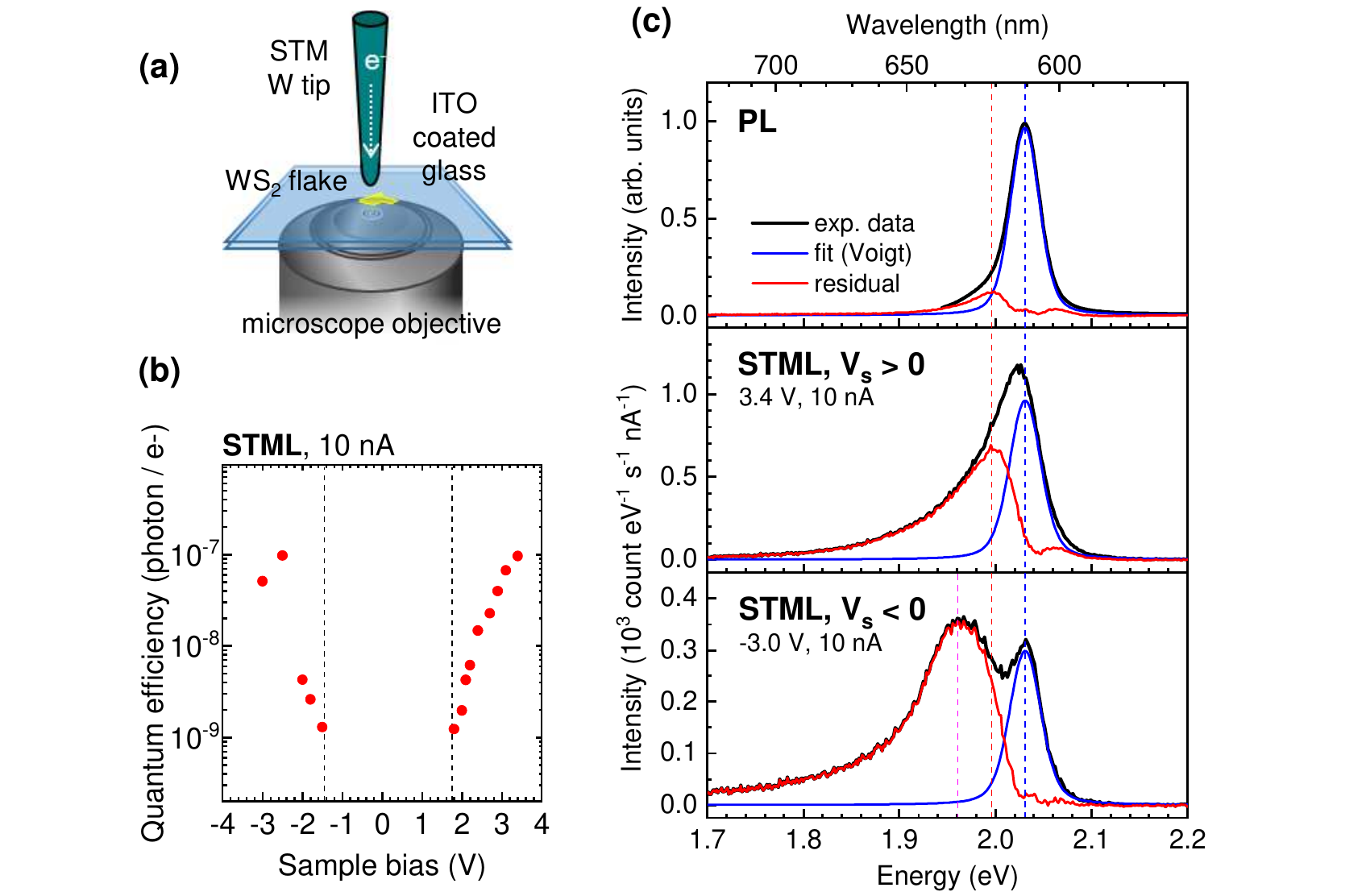} 
	\caption{(a) Schematics of the experiment: the sample is a \ce{WS_2} flake on an ITO-coated glass coverslip, which is placed between the tungsten tip of an STM and a high-NA microscope objective. (b) STML quantum efficiency versus sample bias on a semilog scale, obtained by integrating STML spectra over the $[560,800]$~nm wavelength range. The detection efficiency of the optical setup (available in the Supplemental Material) is taken into account. All STML measurements are carried out at a current setpoint of $10$~nA. (c) PL and STML spectra of monolayer \ce{WS_2}. PL: laser excitation at $\lambda=465.8$~nm, longpass filter from $\lambda=491$~nm. STML: sample bias $V_s=3.4$~V or $-3.0$~V, setpoint current $I_t=10$~nA, acquisition time $t=150$~s. All spectra are fitted using a Voigt profile (blue line) with the same energy position ($2.031$~eV) and similar widths (PL: $35$~meV, STML: $39$~meV) and the residual of the fit is plotted in red. Vertical dotted lines are shifted by $0$ (blue), $-35$ (red) and $-70$~meV (magenta) with respect to the energy position of the neutral A exciton ($2.031$~eV), which is inferred from the fit of the PL peak. 
	}
	\label{FIG-1}
\end{figure*}

Figure~\ref{FIG-1}(a) shows the experimental setup and the principle of the experiment. An air-operated STM head is mounted on top of an inverted optical microscope~\cite{Wang2011,LeMoal2013,Cao2017}. Exfoliated \ce{WS_2} microflakes are deposited onto an indium tin oxide (ITO)-coated glass coverslip using a dry transfer method~\cite{Castellanos-Gomez2014} (ITO thickness $85$~nm). The STM tip is an electrochemically etched tungsten wire. PL is excited using a linearly polarized continuous-wave argon-ion laser emitting at a wavelength of $465.8$~nm, under wide-field illumination in normal incidence. The emitted light is detected in transmission through the substrate using a high-numerical-aperture (NA $=1.49$), oil-immersion microscope objective. The angular distribution of the emitted light is measured by recording an image of the back focal plane of the objective on a CCD camera, i.e., via Fourier-space optical microscopy. (A real-space optical microscopy image is also shown in the Supplemental Material.) All spectra are corrected for detection efficiency and expressed in counts per time, electric current and energy units. 

\section{Results and discussion \label{Resul}} 

Figure~\ref{FIG-1}(b) shows the bias-dependence of the STML quantum efficiency, i.e., the number of emitted photons (in the collection solid angle) per tunneling electron, which is obtained by integrating STML spectra measured on monolayer \ce{WS_2} at a current setpoint of $10$~nA. Light is detected at both bias polarities. Here, the highest quantum efficiencies measured at positive and negative sample bias have the same order of magnitude, i.e., $10^{-7}$~photons per tunneling electron. In Fig.~\ref{FIG-1}(b), the emission onset is observed at a voltage value of about $1.8$~V at positive sample bias and $-1.5$~V at negative sample bias. Below these voltage thresholds, i.e., in the range between the two dotted lines in Fig.~\ref{FIG-1}(b), emission from monolayer \ce{WS_2} cannot be distinguished from the noise background in the measured STML spectra.

In Figure~\ref{FIG-1}(c), we compare two STML spectra for a \ce{WS_2} monolayer measured at sample biases of opposite signs and a PL spectrum from the same sample. The PL spectrum features a peak at $2.031$~eV, which we assign to the radiative recombination of the neutral A exciton ($X$)~\cite{Wang2017a,Lorchat2020}. The PL spectrum fits a Voigt profile of full-width-at-half-maximum (FWHM) $36$~meV centered on $2.031$~eV well, except for the slight peak asymmetry revealed by the residual of the fit. Such an asymmetry may result from phonon coupling or a minor contribution from charged excitons, i.e., trions. According to previous PL studies, the intravalley singlet ($T_S$) and intervalley triplet ($T_T$) trion energies are about $42$ and $35$~meV lower than that of the neutral exciton, respectively~\cite{Vaclavkova2018,Jadczak2019,Robert2021}. Vertical dotted lines at $2.031$~eV (blue line) and redshifted by $35$~meV (red line) are added in Fig.~\ref{FIG-1}(c) as a guide for the eye. 

The STML spectrum at positive sample bias ($V_s=3.4$~V) exhibits a single intensity maximum at $2.026$~eV and its FWHM ($68$~meV) is about twice as large as that of the PL peak. We assume that the STML spectrum is the sum of several contributions centered at different energies, where the highest-energy contribution is that of the neutral A exciton ($X$). Thus, we fit the high-energy side of the STML spectrum using a similar Voigt profile as for the PL peak (same energy $2.031$~eV, FWHM of $39$~meV). The residual of the fit reveals an asymmetric peak with an intensity maximum redshifted by about $35$~meV with respect to the higher-energy peak, with a exponential decay tail on the low-energy side (more details are available in the Supplemental Material). Such an exponential tail has been previously reported for trion PL peaks and ascribed to the electron recoil effect~\cite{Esser2001,Ross2013}. The area under the curve of the fit residual is about $60\%$ of the total emitted light. Based on this simple analysis, we infer that STML from monolayer \ce{WS_2} at $V_s>0$ primarily consists of the radiative recombination of charged and neutral A excitons, with a slightly larger ($60\%$) trionic contribution. 

%
\begin{figure*}
	\includegraphics{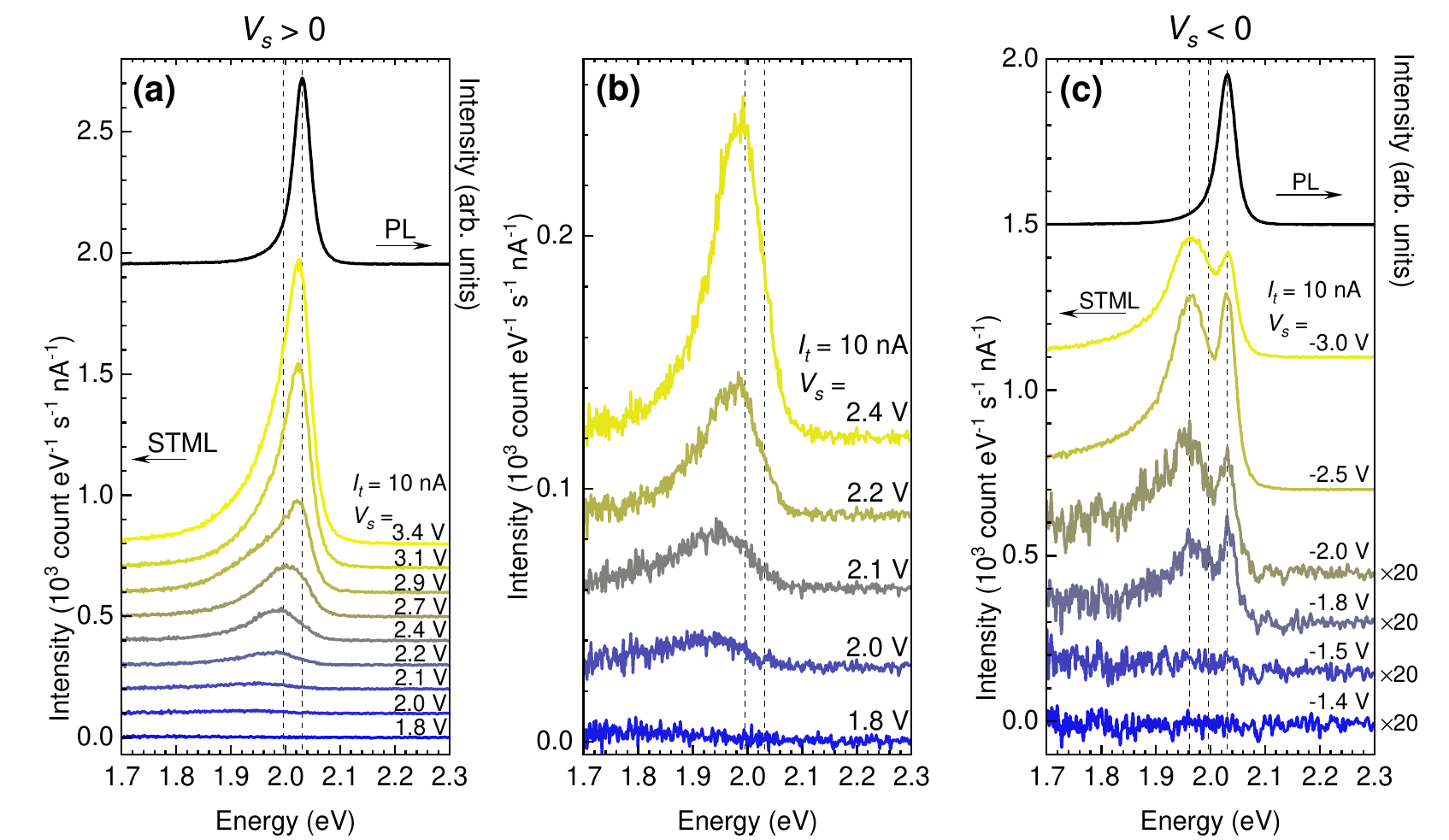} 
	\caption{STML spectra versus sample bias and PL spectra of monolayer \ce{WS_2}. (a) Positive sample bias from $1.8$ to $3.4$~V (from bottom to top); (b) same data as shown in panel a, from $1.8$ to $2.4$~V, with a magnified intensity scale. (c) Negative sample bias from $-1.4$ to $3.0$~V; intensity is multiplied by $20$ for the data measured at biases from $-1.4$ to $-2.0$~V. PL (black line, right scale): laser excitation at $\lambda=465.8$~nm, longpass filter from $\lambda=491$~nm. STML (colored lines, left scale): $I_t=10$~nA, $t=150$~s. Vertical dotted lines shifted by $0$, $-35$ and $-70$~meV with respect to the neutral A exciton energy ($2.031$~eV), as inferred from PL, are shown as a guide for the eye.
	}
	\label{FIG-2}
\end{figure*}

The STML spectrum at negative sample bias ($V_s=-3.0$~V) features two local intensity maxima, i.e., at $1.967$ and $2.032$~eV. We fit the higher-energy peak using the same Voigt profile (same energy $2.031$~eV, same FWHM $39$~meV) as for the STML spectrum recorded at $V_s>0$. The residual of the fit is an asymmetrical peak, the intensity maximum of which is redshifted by about $70$~meV with respect to the higher-energy peak. The area under the curve of the fit residual is about $75\%$ of the total emitted light. While a quantitative analysis is difficult, from this simple analysis, we can infer that the STML of monolayer \ce{WS_2} at $V_s<0$ features contributions from excitonic species other than $X$ and $T_S/T_T$, and that the neutral A exciton is excited at both bias polarities. According to previous PL studies~\cite{Jadczak2019,Zinkiewicz2021}, several different excitonic complexes in monolayer \ce{WS_2} have binding energies in the range of $50$ to $70$~meV, including intravalley spin-forbidden and intervalley momentum-forbidden trions and negatively charged biexcitons. In addition, the energy position of the lower-energy peak is also consistent with the luminescence of localized excitonic states, which have been previously observed in monolayer \ce{WS_2}~\cite{Vaclavkova2018}.
 
\begin{figure}
	\includegraphics[width=1.0\linewidth]{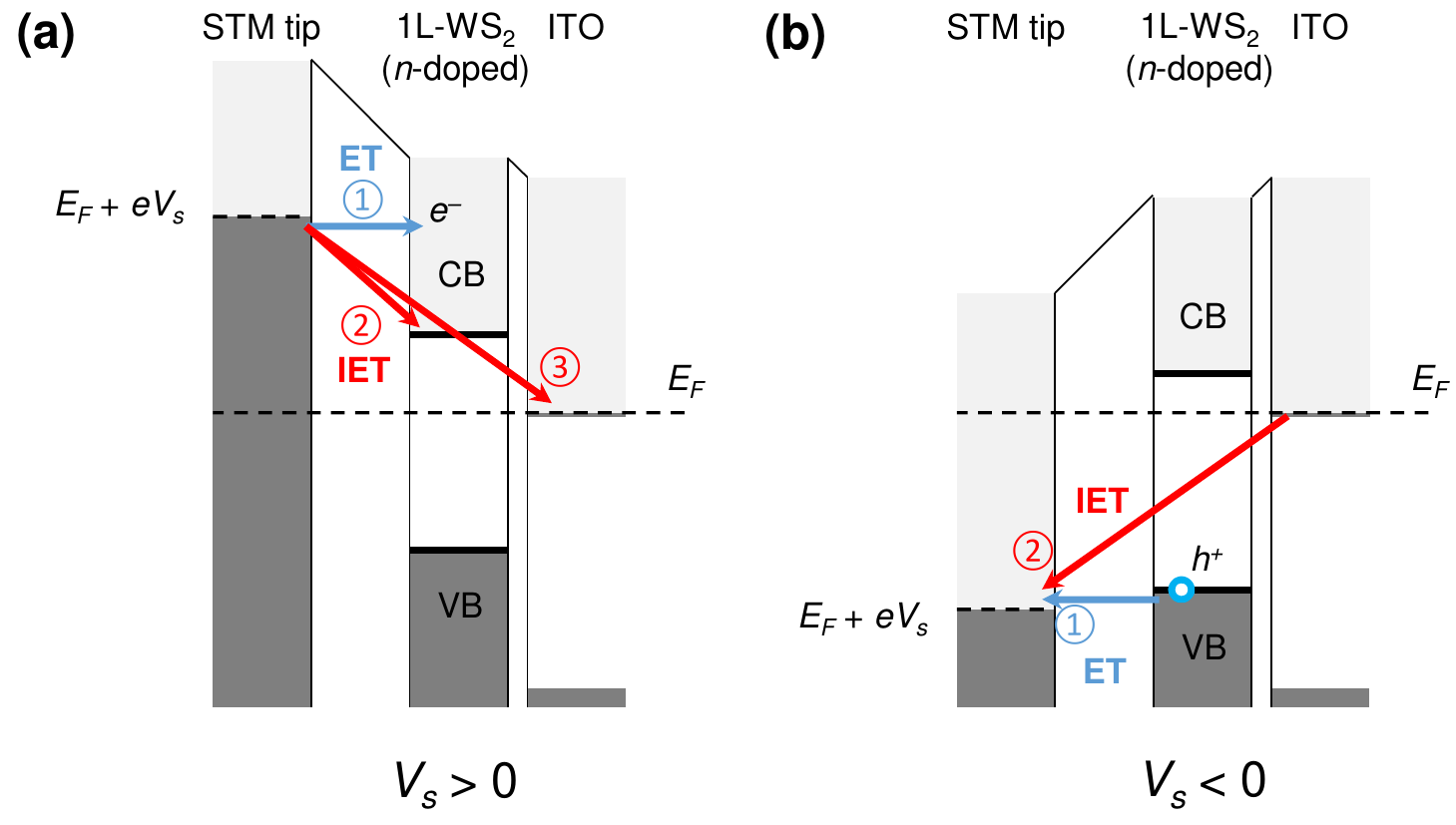}
	\caption{Simplified energy diagrams of the studied system, showing elementary elastic (ET) and inelastic (IET) electron tunneling processes that may occur upon electrically biasing the tip-sample tunnel junction. CB/VB: conduction/valence bands. $E_F$: Fermi level, $V_s$: sample bias voltage, $e$: elementary charge, $e^-$: electron, $h^+$: hole. (a) At positive sample bias, electrons may tunnel elastically (\textcircled{1}) or inelastically (\textcircled{2}) to the conduction band of \ce{WS_2} or to unoccupied surface states of ITO, or to defect or interface-induced gap states in \ce{WS_2} (\textcircled{3}). (b) At negative sample bias, electrons may tunnel elastically (\textcircled{1}) from the valence band of \ce{WS_2} to the tip, thus creating holes in the valence band, or inelastically (\textcircled{2}) from ITO to the tip. At both bias polarities, IET-induced energy transfer is one possible mechanism for the creation of excitons in monolayer \ce{WS_2}. At negative sample bias, excitons may also form from holes created in the valence band and electrons already present in the conduction band of the \textit{n}-doped \ce{WS_2} monolayer. Note that the arrows are very simplistic representations of the electronic processes.
	}
	\label{FIG-3}
\end{figure}

Below, we examine the dependence of the STML spectra on the bias voltage. Figures~\ref{FIG-2}(a) to \ref{FIG-2}(c) show the STML spectra used to calculate the quantum efficiencies given in Fig.~\ref{FIG-1}(b). PL spectra measured on the same sample are added in Figs.~\ref{FIG-2}(a) and \ref{FIG-2}(c). The data shown in Fig.~\ref{FIG-2}(b) are the same as those shown in the bottom part of Fig.~\ref{FIG-2}(a), using a different intensity scale. Moreover, simplified energy diagrams of the studied system are shown in Figure~\ref{FIG-3} to aid in the discussion on the possible excitation mechanisms. 

%

First, we consider the STML data measured at $V_s>0$ shown in Fig.~\ref{FIG-2}(a). We observe a bias-dependent spectral shift, which is most clearly visible at low voltage [see Fig.~\ref{FIG-2}(b)], and a clear effect of the quantum cutoff, where no light is detected at a photon energy higher than $e|V_s|$. The onset of the excitonic STML is between $2.0$ and $2.1$~V, i.e., when $e|V_s|$ is at least equal to the exciton energy (i.e., the optical gap). This observation suggests that non-radiative energy transfer from the tunnel current to the semiconductor~\cite{Pommier2019} is the mechanism underlying the onset of the excitonic STML at $V_s>0$. Such energy transfer may occur through inelastic electron tunneling from the tip to the \ce{WS_2} monolayer or to the substrate, i.e., via channels 2 and 3 in Fig.~\ref{FIG-3}(a). Within this interpretation, we may ascribe the bias-dependent spectral shift observed at low voltage to an effect of the quantum cutoff. Up to $V_s=2.0$~V, the emitted light is almost entirely related to lower-energy (possibly localized~\cite{Vaclavkova2018}) excitonic species or other radiative processes (e.g., radiative electronic transitions between tip and sample states~\cite{Schuler2020}). Above $V_s=2.0$~V, the relative contribution of trions and excitons to the total radiated power increases with the bias voltage and, beyond a certain voltage [about $2.9$~V in Fig.~\ref{FIG-2}(a)], bias-dependent spectral shifts are negligible.

Unlike what is measured at $V_s>0$, the data recorded at $V_s<0$ do not exhibit a significant bias-dependent spectral shift [see Fig.~\ref{FIG-2}(c)]. From $V_s=-3.0$ to $-1.8$~V, all spectra feature the same two characteristic peaks, i.e., a higher-energy peak centered on the same energy position as the PL peak ($2.032$~eV) and a comparatively broader and more asymmetric peak shifted by about $70$~meV as compared to the PL (i.e., at $1.967$~eV). The intensity ratio of the two peaks, which slightly varies from one spectrum to another in Fig.~\ref{FIG-2}(c), does not depend on the bias voltage. These two peaks are not present (or cannot be distinguished from the background) in the spectrum recorded at $V_s=-1.5$~V; thus, we estimate that the onset of the excitonic STML is between $-1.8$ and $-1.5$~V. At $V_s=-1.8$~V, overbias emission occurs at a photon energy about $0.2$~eV higher than the quantum cutoff $e|V_s|=1.8$~eV, with clear evidence of neutral A exciton excitation. All together, these observations are consistent with an STML excitation mechanism based on hole injection in the valence band of an $n$-doped semiconductor, which is described as channel~1 in Fig.~\ref{FIG-3}(b). Indeed, if the Fermi level of the tip is lower in energy than the valence band maximum of the \ce{WS_2} monolayer, electron tunneling from the sample to the tip may create holes in the valence band of the semiconductor. These holes may bind with electrons already present in the conduction band to form excitons~\cite{Reinhardt2010} (assuming that the \ce{WS_2} monolayer is \textit{n}-doped). Nevertheless, such a mechanism may coexist with or be superseded by energy transfer~\cite{Pommier2019} at $V_s<-2.0$~V [see channel~2 in Fig.~\ref{FIG-3}(b)]. Foremost, the absence of a spectral shift in the STML data recorded at different bias voltages tends to rule out the possibility that the lower-energy peak results from redshifted contributions of $X$ and $T_S/T_T$, e.g., due to the static electric field in the tunneling junction (i.e., Stark effect).

%
\begin{figure*}
	\includegraphics{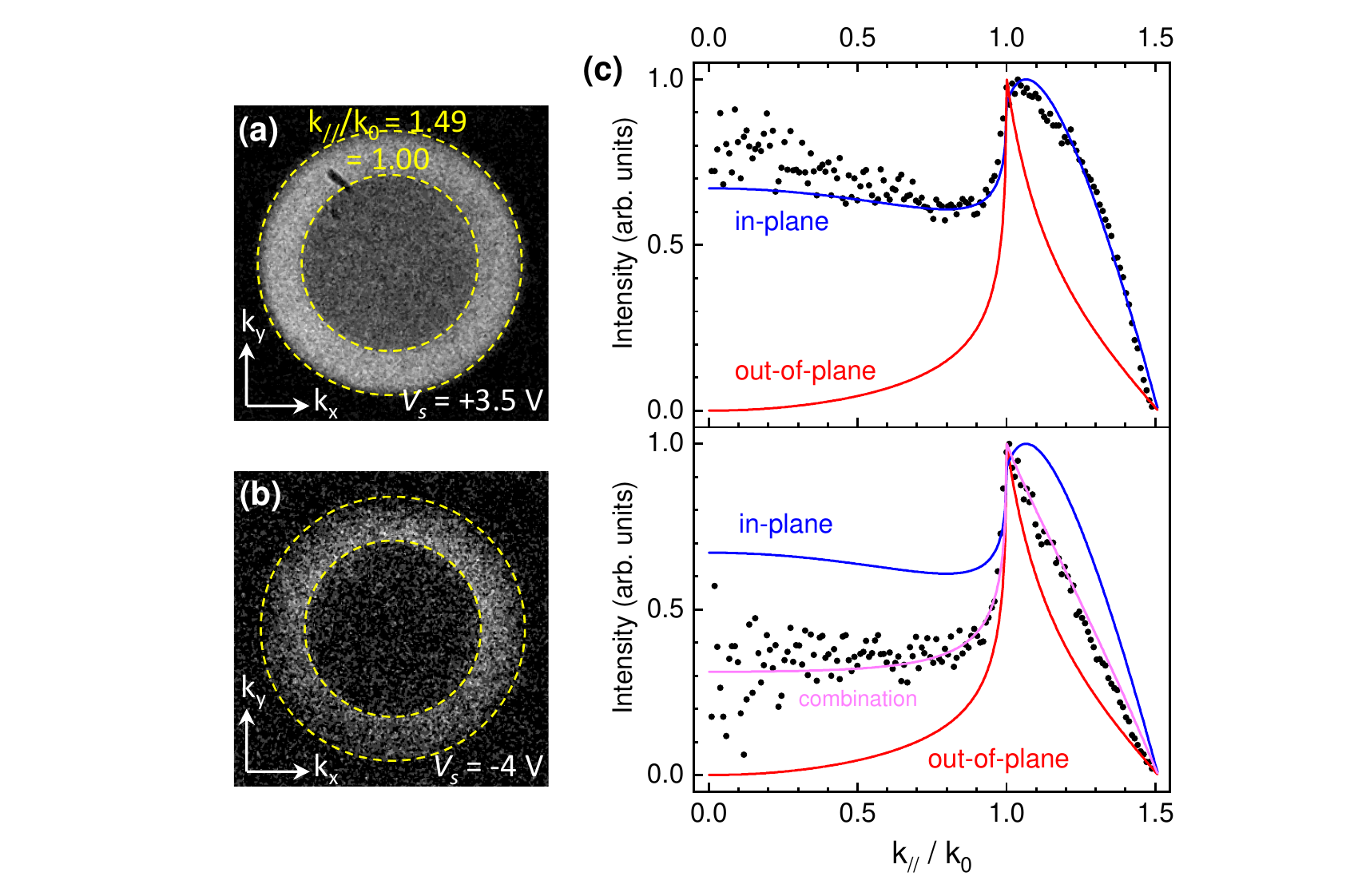} 
	\caption{Bias polarity dependence of the STML radiation pattern. [(a) and (b)] Fourier-space optical microscopy images of the STML measured on monolayer \ce{WS_2} at sample bias (a) $V_s=3.5$~V and (b) $V_s=-4.0$~V ($I_t=10$~nA). The inner and outer circles (yellow dashed lines) indicate the critical angle of the air/glass interface and the largest collection angle, respectively. (c) Intensity profiles averaged over all directions in Fourier space (black dots), taken from the images shown in a (top) and b (bottom), and simulations using the model of an oscillating electric dipole on an air-glass interface. In the model, the dipole is oriented in-plane (blue curve) or out-of-plane (red curve) with respect to the interface; moreover, we show a linear $0.45/0.55$ combination of the two simulated curves (magenta curve). The experimental profiles are corrected for apodization~\cite{Kurvits2015}, i.e., multiplied by $\cos\theta$ where $\theta$ is the collection angle. $k_{\parallel}/k_0=n\sin{\theta}$ is the normalized in-plane coordinate in Fourier space, where $k_0=\omega/c$ is the photon wavevector modulus and $n$ is the refractive index of glass. Experimental data and simulated curves are normalized to unity.
	}
	\label{FIG-4}
\end{figure*}

Finally, we investigate the bias polarity-dependence of the STML radiation pattern. Figures~\ref{FIG-4}(a) and \ref{FIG-4}(b) show Fourier-space optical microscopy images~\cite{Cao2017} of the angular distribution of the STML measured at positive ($V_s=3.5$~V) and negative ($V_s=-4.0$~V) sample bias, respectively. Intensity profiles from these images, averaged over all in-plane directions, are plotted in Fig.~\ref{FIG-4}(c) and compared to simulations based on a simple analytical model. In the model, we consider a point-like oscillating electric dipole on an air/glass interface, which is either oriented parallel (i.e., in-plane) or perpendicular (i.e., out-of-plane) to the interface and we calculate the far-field angular distribution of the radiated power in glass~\cite{Pommier2019}. At $V_s>0$, the simulation of an in-plane electric dipole reproduces the experimental data well. Such good agreement confirms that the STML of monolayer \ce{WS_2} at $V_s>0$ results from radiative recombination of spin-bright excitonic species (i.e., from the spin-allowed optical transitions), whose transition dipole moment is in the plane of the monolayer. This result is similar to our previous observations for the STML of monolayer \ce{MoSe_2}~\cite{Pommier2019}; however, the result is strikingly different for monolayer \ce{WS_2} at $V_s<0$. In this case, the best fit is obtained using a linear combination of the two simulated curves, i.e., an incoherent superposition of the radiation patterns calculated for the in-plane and out-of-plane electric dipoles. As shown in Fig.~\ref{FIG-4}(c), we use a $0.45/0.55$ in-plane/out-of-plane combination of the simulated curves normalized to unity (or $0.78/0.22$ of the as-calculated curves), which corresponds to $25\%$ ($75\%$) of the detected light being emitted by out-of-plane (in-plane) electric dipoles. This is an approximation, since we neglect the frequency-dependence of the detection efficiency in the experiment. Nevertheless, this result indicates that the low-energy peak in the STML spectra measured at $V_s<0$ [see Figs.~\ref{FIG-1}(c) and~\ref{FIG-2}(c)] is not purely out-of-plane electric dipoles, since the area under the low-energy peak is about three times as large as that behind the high-energy peak, which we assign to excitons that have in-plane transition dipole moment. Therefore, the low-energy peak itself must result from several contributions, only a minor part of which corresponds to out-of-plane dipole emission. 

The observation of out-of-plane dipole contributions in STML at (and only at) negative sample bias may be interpreted in several different ways. On the one hand, part of the emitted light may not be of excitonic origin. Based only on the fit shown in Fig.~\ref{FIG-1}(c), we cannot exclude the presence of a spectrally broad contribution (of non-excitonic origin) in addition to the two STML peaks that we ascribe to excitonic luminescence at $V_s<0$. A spectrally broad contribution of this type could explain the stronger and longer low-energy tail observed at $V_s<0$ as compared to $V_s>0$ [see the $1.7-1.9$~eV energy part of the spectra shown in Fig.~\ref{FIG-1}(c) and Fig.~\ref{FIG-2}]. Indeed, radiative inelastic electron tunneling between the tip and the monolayer may also occur without the creation of excitons~\cite{Schuler2020}. The transition dipole moment associated with such a radiative process is oriented in the tunneling current direction~\cite{Johansson1998}, i.e., perpendicular to the sample surface. This non-excitonic emission is expected to be spectrally broad, because the STML spectrum of the empty tunneling junction, on ITO in absence of a TMD flake, is featureless and spans an energy range from the near-infrared to the quantum cut-off~\cite{Pommier2019} (see the spectrum in the Supplemental Material). Thus, non-excitonic emission could contribute to the low-energy background observed in STML spectra at $V_s<0$. However, Fourier-space images systematically feature an out-of-plane dipole contribution at $V_s<0$, which we never observe at $V_s>0$. Such a bias polarity dependence is not expected for this non-excitonic emission, since such a phenomenon is based on energy transfer from the tunnel current to photonic modes available in the junction, the density of which is independent of the current direction. 

On the other hand, the STML of monolayer \ce{WS_2} at $V_s<0$ may not only result from the radiative decay of spin-bright, but also spin-dark excitonic species (i.e., from spin-forbidden optical transitions), which may have a non-zero transition dipole moment perpendicular to the monolayer. Tip-enhanced PL microscopy of spin-dark excitons has been previously reported~\cite{Park2018}. The use of a plasmonic tip or nanocavity, e.g., a gold tip on a gold surface, is required, in order that the local electric field enhancement compensates for the radiative quantum yield of dark excitons, which is orders of magnitude lower than those of spin-bright excitons. Here, this effect is not present, since ITO and tungsten do not support surface plasmons in the investigated frequency range~\cite{LeMoal2016}, whatever the bias polarity. 

Moreover, tip polarization effects may be responsible for out-of-plane dipole contributions. Even spin-bright excitonic species, which have an in-plane transition dipole, may induce an image dipole in the tip that has both in-plane and out-of-plane components, provided that the excitonic species are located off the tip axis. The reason why such an effect depends on bias polarity may reside in the in-plane spatial distribution of emitting dipoles around the tip, which depends in turn on the diffusion length of the excited excitonic species. Indeed, the strength and the orientation of the image dipole in the tip is determined by the exciton-tip distance. Moreover, some of the excited excitonic species may be localized or trapped in or repeled from the junction due to the strong static electric field~\cite{Unuchek2018}.  

\section{Conclusions \label{Concl}} 

To conclude, we have investigated the STML of monolayer \ce{WS_2} on ITO-coated glass in air using an STM coupled to an optical microscope. Thus, we have shown that the bias polarity of the tip-sample junction has an effect on both the spectral and radial distribution of the emitted light. At $V_s>0$, the dominant contribution to the STML of monolayer \ce{WS_2} is from the radiative recombination of charged (trion) and neutral A excitons, whereas at $V_s<0$, other (lower-energy) excitonic species also play a role. The STML excitation of the neutral A exciton occurs for both bias polarites, and the resulting emisson energy is similar to that measured using laser-induced PL spectroscopy. At $V_s>0$, the STML radiation pattern matches that of an in-plane transition dipole. This demonstrates that only radiative recombinations of spin-bright excitonic species are involved and also rules out contributions from non-excitonic processes. These conclusions are similar to those found for monolayer \ce{MoSe_2}~\cite{Pommier2019}. At $V_s<0$, the STML radiation pattern results from a linear combination of in-plane and out-of-plane electric dipoles, with a dominant contribution of in-plane dipoles. Such an out-of-plane dipole contribution, which is only observed at $V_s<0$, could result from the excitation of spin-dark excitons or inelastic tunneling-induced light without the creation of excitons; however, such processes are not expected to depend on the bias polarity. Nevertheless, tip polarization effects may yield an out-of-plane dipole contribution, even if only excitons with an in-plane transition dipole moment are excited. Such effects are determined by the spatial distribution of the excitons around the tip position. This distribution may depend on the bias polarity, due to the effect of the static electric field on the exciton diffusion. Overall, the combination of STML spectroscopy and wide-field optical microscopy reveals crucial information that is not accessible in a conventional STML imaging mode, i.e., via tip-scanning photon mapping. In particular, the orientation of the transition dipole moment of the emitters may be retrieved from their angular radiation pattern, which is recorded using Fourier-space optical microscopy. In principle, such information may be mapped on the nanometer scale, by simultaneously recording Fourier-space optical microscopy images and scanning the tip. This opens up new prospects, e.g., for the nanoscale study of emergent excitonic physics in TMD van-der-Waals heterostructures~\cite{Wilson2021}. Finally, we suggest that more insight into the STML mechanisms of TMD monolayers and heterostructures may be obtained from STML experiments on gated TMD samples, where electrostatic tuning of the charge carrier density is possible.

\begin{acknowledgments}
This work is supported by public grants from the French National Research Agency (H2DH ANR-15-CE24-0016, ATOEMS ANR-20-CE24-0010, Intelplan ANR-15-CE24-0020 and M-Exc-ICO ANR-16-CE24-0003) and overseen by the ANR as part of the ``Investissements d'Avenir'' program (Labex NIE ANR-11-LABX-0058-NIE and Labex NanoSaclay ANR-10-LABX-0035). We acknowlege the StNano clean room staff for technical support. S.B. acknowledges support from the Indo-French Centre for the Promotion of Advanced Research (CEFIPRA) and from the Institut Universitaire de France (IUF). This project has received funding from the European Research Council (ERC) under the European Union's Horizon 2020 research and innovation program (grant agreement No 771850). This work of the Interdisciplinary Thematic Institute QMat, as part of the ITI 2021-2028 program of the University of Strasbourg, CNRS and Inserm, was supported by IdEx Unistra (ANR 10 IDEX 0002), and by SFRI STRAT'US project (ANR 20 SFRI 0012) and EUR QMAT ANR-17-EURE-0024 under the framework of the French Investments for the Future Program. This work has received financial support from the Funda\c{c}\~{a}o de Amparo \`{a} Pesquisa do Estado de S\~{a}o Paulo (FAPESP), through projects 18/08543-7, 20/12480-0 and 14/23399-9.
\end{acknowledgments}

\end{document}